\pdfoutput=1
\documentclass[cits]{JINST}
\let\ifpdf\relax
\usepackage{upgreek} % Needed for the non-italicized mu for micrometers.
\usepackage{url}
\usepackage{multirow}
\usepackage{lineno}

\providecommand{\etal}{~et~al.\ }

\title{Enhancement of NEST Capabilities for Simulating Low-Energy Recoils in Liquid Xenon}

\author{M. Szydagis$^a$\thanks{Corresponding author.}, A. Fyhrie$^b$, D. Thorngren$^a$, M. Tripathi$^a$\\
\llap{$^a$}Physics Department, University of California, Davis,\\
 1 Shields Avenue, Davis, CA 95616, USA\\
\llap{$^b$}Physics Department, University of California, Berkeley,\\
 101 Sproul Hall, Berkeley, CA 94704, USA\\
 E-mail: \email{mmszydagis@ucdavis.edu}}

\abstract{The Noble Element Simulation Technique (NEST) is an exhaustive collection of models explaining both the scintillation light and ionization yields of noble elements as a function of particle type (nuclear recoil, electron recoil, alphas), electric field, and incident energy or energy loss ($dE/dx$). It is packaged as C++ code for Geant4 that implements said models, overriding the default model which does not account for certain complexities, such as the reduction in yields for nuclear recoils (NR) compared to electron recoils (ER). We present here improvements to the existing NEST models and updates to the code which make the package even more realistic and turn it into a more full-fledged Monte Carlo simulation. All available liquid xenon data on NR and ER to date have been taken into consideration in arriving at the current models. Furthermore, NEST addresses the question of the magnitude of the light and charge yields of nuclear recoils, including their electric field dependence, thereby shedding light on the possibility of detection or exclusion of a low-mass dark matter WIMP by liquid xenon detectors.}


\begin{document}

\section{Motivation and Description} %1
\label{sec:intro} % 1

Liquid and gaseous noble elements have been established as versatile target materials in a wide variety of detectors in physics experiments and also in a broader context of particle detection. They have been utilized in experimental searches for WIMP dark matter, for detection of neutrinoless double beta decay, for reactor monitoring via coherent neutrino scattering, for long-baseline neutrino experiments, in rare decay high-energy physics experiments, and in medical physics for PET scans~\cite{AraujoChepel}. The original set of NEST models~\cite{Szydagis} was arrived at by compiling all measurements in the vast literature on noble elements, especially liquid xenon, and combining everything learned into a unified approach. Here we again focus on liquid xenon, but present updates made to the NEST framework since its original inception. An examination of~\cite{Szydagis}, and the many references therein representing over forty years of measurements of xenon, will lay the foundation for better understanding the results presented in this work. 

The NEST code repository is available for public download at~\cite{NestSite}, and is utilized by the communities involved in noble liquid (or gas) based technologies, in either the single or dual-phase, or zero or non-zero electric field configurations, the latter being time-projection chambers (TPCs). NEST replaces the existing physics process within Geant4~\cite{Agostinelli2003,Geant4} for providing scintillation photons (G4Scintillation) and treats scintillation from excitation and recombination separately~\cite{Szydagis,NestSite}. In its implementation it introduces a new particle type, the thermal electron, distinct from electrons (analogous to the difference between gammas and optical photons), which enables the simulation of electrons drifting in a field after ionization. It also adds a new process to simulate the electroluminescence signal in TPCs~\cite{Mock2013}. NEST differentiates itself from most earlier simulation efforts by directly predicting absolute numbers of photons and electrons produced by an energy deposition~\cite{AraujoChepel,Szydagis}, a concept fully developed by Dahl~\cite{Dahl2009}. This approach is more powerful as compared to the earlier practice of only quoting relative yield, for instance in NR measurements. It also suggests a superior technique for calorimetry in xenon TPCs (by treating scintillation and ionization as being equally important), which has thus far not been fully employed in two-phase xenon-based dark matter experiments, XENON10/100 and ZEPLIN~\cite{Aprile2010a,Aprile2010,Araujo2006}.

\section{ER Yields vs. Energy and Field} %2
\label{sec:keVee} % 2

We begin by discussing the light and charge yields of electron recoils per unit energy, from gamma rays or beta particles. They are non-linear with respect to energy and electric field strength. This issue was already treated extensively in~\cite{Szydagis}. However, the conventional thought continues to persist in the field (for instance, see~\cite{McKLidTalk}) that the absolute light yield for electron recoils in liquid xenon is $\sim$40~photons/keV at zero field (comparable to liquid argon), a value stemming from seminal studies on absolute yield conducted by Doke\etal\cite{Doke1988,Doke2002}. This number applies only to electrons not gammas, and only at 1 MeV, so it is not applicable to dark matter experiments. As the yield increases with increasing $dE/dx$ (generally decreasing energy) due to increasing recombination, it can be as large as 60 photons/keV or even higher, at energies $O$(10-100) keV. The evidence for this claim is both indirect and direct, as discussed below.

The indirect evidence~\cite{Yamashita2003,Barabanov1987} is gleaned from measuring light yield at $\sim$1 MeV and at different, lower energies with the same detector, and observing a strong relative increase in yield. The direct evidence is provided by experiments where the light collection and photon detector efficiencies were estimated with Monte Carlo methods and used to extract absolute photon yields for energies below 1 MeV~\cite{Ni2006,Shutt2007}. Further direct evidence can be extracted from Plante\etal\cite{Plante2011}, who report 24.14$\pm$0.09(stat)$\pm$0.44(sys) photoelectrons/keV for $^{57}$Co. Assuming a high photomultiplier tube (PMT) quantum efficiency of 40\% and 100\% light collection efficiency because of the nature of their detector (a cube where PMTs are the sides) this corresponds to at least 60 photons/keV at $\sim$122 keV. An understanding of the difference between photons/photoelectrons and light yield/light collection helps interpret reported values of light yield, for example, that of the XENON100 detector, quoted as 2.28$\pm$0.04 photoelectrons/keV for $^{57}$Co~\cite{DarkAttack}. All values like this from real detectors appear low because one must take into account the photon detection efficiency, which is the product of the light collection (never perfect in large-scale detectors because of non-unity reflectivities and finite photon absorption lengths) and quantum efficiency, and the electric field, which reduces the yield with respect to zero field by quenching the recombination process.

Given new results on low-energy ER provided by Compton Scattering experiments~\cite{Baudis2013,Lim2013}, contradicting previous work in the same energy range~\cite{Obodovskii1994}, the original NEST zero-field ER model required revision. There was qualitative agreement between simulation and data: when the energy was decreased from 1 MeV, a point was reached at $O$(10) keV where the increase in light yield per unit energy reached a maximum and started decreasing. However, the Thomas-Imel box model~\cite{ThomasImel} recombination parameter of 0.19 used in~\cite{Szydagis}, based solely on the only available data set at the time~\cite{Obodovskii1994}, overestimated the absolute yield by $\sim$20\%. Instead of applying an ad hoc correction, we sought a microphysics-based reason for a lower value, and discovered the following. An insertion of 2 V/cm, which is the approximate value of the (screened) electric field of a xenon nucleus at the electron-ion thermalization radius of 4600 nm~\cite{Mozumder1995}, into the power law fit for the field dependence of the Thomas-Imel recombination factor (which has an exponent of -0.1, physically motivated and based on~\cite{Dahl2009}) yields a recombination parameter of 0.05. This agrees with the non-zero field values for this factor from Dahl ($\sim$0.02-0.04~\cite{Dahl2009}) much more closely than the original 0.19, thus smoothing the transition from zero to non-zero field in NEST. It also appears to be more physical because it incorporates the fact that even in the zero-field case there is still the inherent field of the xenon nucleus that can affect the recombination probability.

\begin{figure}
\begin{centering}
\includegraphics[width=0.85\textwidth]{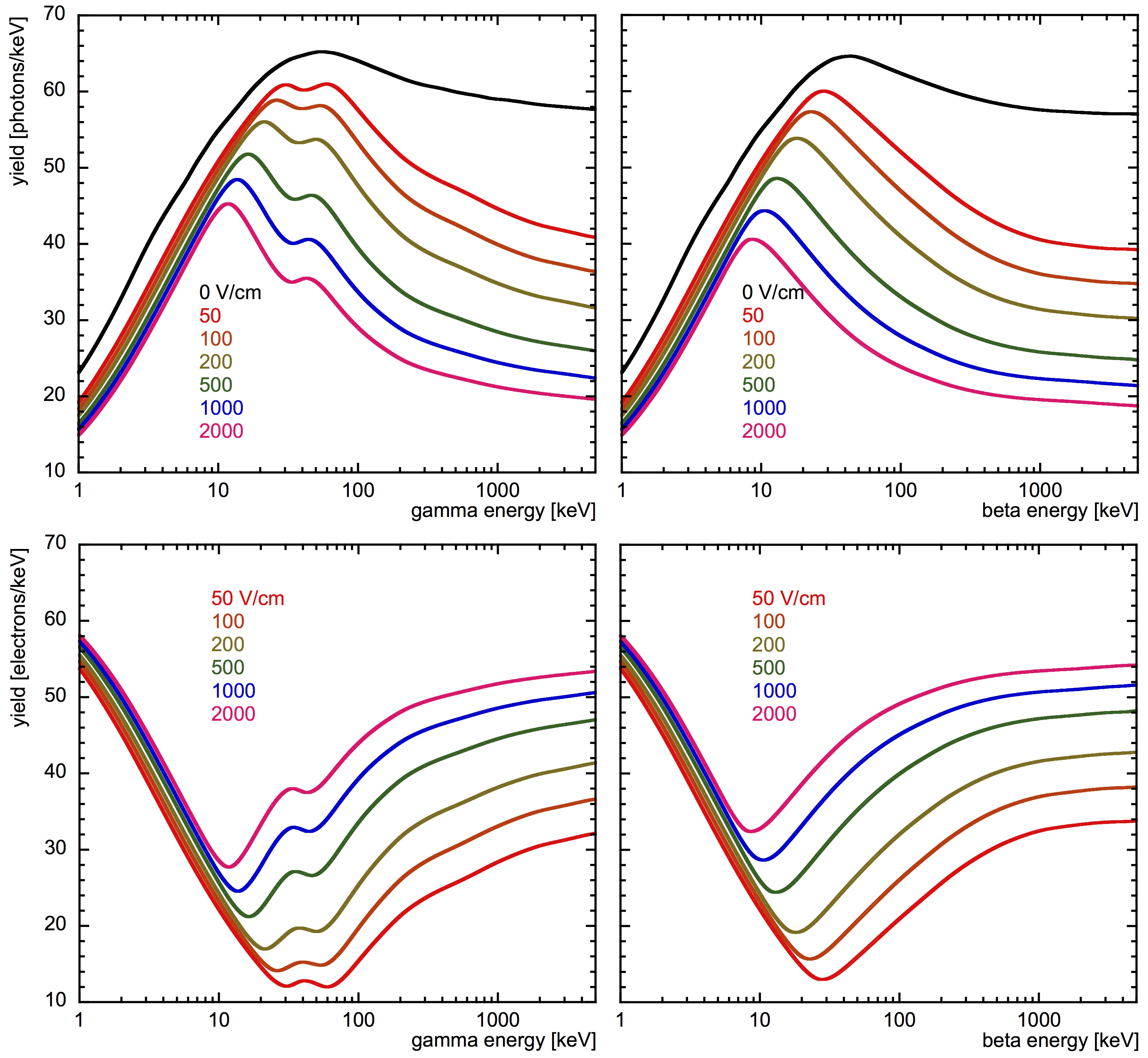}
\caption[]{Predictions from NEST for the light (photons/keV, top row) and charge (electrons/keV, bottom row) yields of electron recoil events when a gamma ray is the primary particle interacting with liquid xenon (left) and when a pure electron, a beta particle or a delta ray, is the primary particle (right). Yields are shown as a function of both the total energy and the electric field. Higher electric field reduces recombination, raising the charge yield at the expense of light, in an anti-correlated fashion. Comparison to data is performed in Figure 7 of~\cite{Szydagis}, which presents the NEST model in an alternative form, plotting parallel curves of yields versus field for different energies. The dip in the gamma curves, analogous to features observed in the scintillation yield versus energy of NaI~\cite{Tawara2000}, is caused by the xenon K-edge x-ray that creates a second possible interaction site, displaced from the site of the initial energy deposition. The lower light yields for electrons are due to their lower $dE/dx$ as compared to gammas of the same initial energies.}
\label{ERFourSquare}
\end{centering}
\end{figure}

Figure~\ref{ERFourSquare} shows the NEST curves from the new approach contained in version 0.98. The non-zero field curves are unchanged (see~\cite{Szydagis} for their comparison to data). To observe the excellent agreement with data from zero field, please see Baudis et al.~\cite{Baudis2013}.

\section{NR Yields vs. Energy and Field} %3
\label{sec:keVnr} %3

In this section, we refine the zero-field NR light yield first reported in~\cite{Szydagis}. Instead of an exciton-to-ion ratio of 0.06 (originally chosen for simplicity, using the same value for ER and NR), we now use 1.00 for NR as an ansatz solution to confront it with available data. We find that it has excellent agreement with the results of~\cite{Plante2011} and~\cite{Horn2011}, well within their 1-sigma experimental error bars~\cite{SzydIDM2012}.

Evidence for a higher ratio is plentiful: (1) NR yields change more slowly with field than ER, implying recombination plays a smaller role~\cite{Dahl2009}, (2) NR recombination fluctuations are smaller~\cite{Dahl2009}, implying the same, (3) when fitting field data, allowing this ratio to be a free parameter one finds best fits near 1~\cite{Dahl2009}, and (4) values near 1 permit usage of nearly the same Thomas-Imel recombination probability parameter as for ER, at any field, described by the power law mentioned in Section~\ref{sec:keVee}. Other works support a ratio at or near 1 with similar~\cite{Sorensen2011} or completely different approaches~\cite{Bezrukov2011}. To the best of our knowledge, however, only NEST provides absolute numbers for the light yield instead of just a recombination fraction~\cite{Sorensen2011}, or the traditional comparison to $^{57}$Co gammas, which is often labeled as relative yield compared to ER~\cite{Aprile2006} even though we know from~\cite{Szydagis} and Figure~\ref{ERFourSquare} that ER scintillation yield versus energy is not constant. This "standard candle" is not even the same for different detectors. It is a composite line of 122 and 136~keV~\cite{LBLTOI}. Depending on detector composition and shielding, the average energy deposited varies, and this low energy is easily stopped by self-shielding, thus making it impossible to use it to calibrate the center of a large-scale detector. This necessitates an extrapolation from small-scale detectors.

\begin{figure}
\begin{centering}
\includegraphics[width=0.99\textwidth]{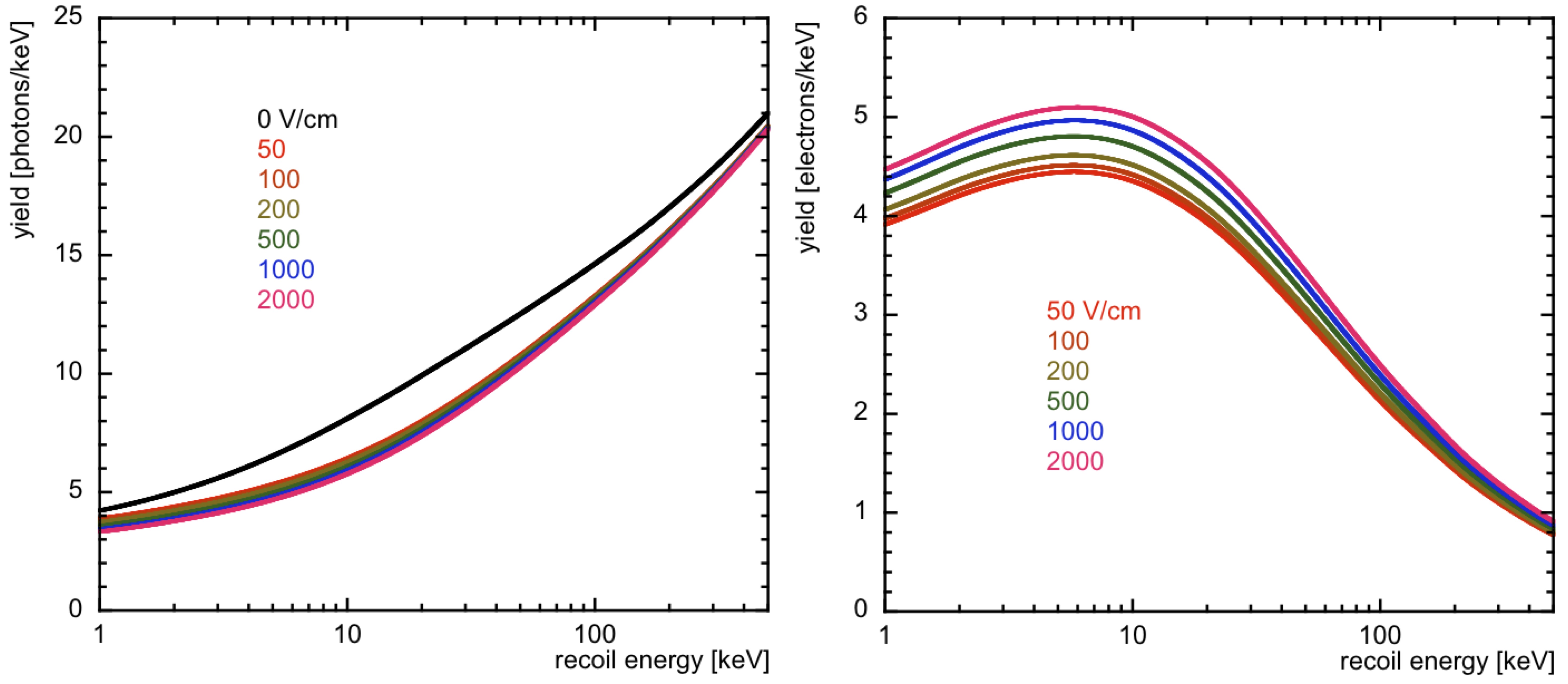}
\caption[]{The light (photons/keV, left) and charge (electrons/keV, right) yields of nuclear recoil events, as a function of both energy and electric field. Comparison with data is not shown in either case. For light, there is no direct comparison since no experiment quotes absolute yield. However, a comparison is possible if an assumption of 63~photons/keV is made for a 122~keV gamma~\cite{SzydIDM2012}. The turn-over in the charge yield curve is caused by the decrease in the total number of quanta (as described by the Lindhard factor) beginning to dominate over the increase in the charge yield resulting from the decreasing Thomas-Imel recombination probability (a smaller total number of ions is being created). Discussion of charge yield in measurements is contained in the text.}
\label{NRTwoSquare}
\end{centering}
\end{figure}

Change in yield with field is reported by Dahl~\cite{Dahl2009}. However, his setup could not record neutron scatter angle, so it could not measure energy directly. An assumption is needed to set an absolute scale. For this we cite Hitachi (Figure 5 in~\cite{Aprile2006}), who performed a first-principles calculation of the NR $dE/dx$. His result can be approximated in an analytical form with the Lindhard parameterization~\cite{Lindhard1963}, by adjusting one value~\cite{Sorensen2011}. We argue that using Dahl's data, despite this disadvantage, is an improvement over the earlier technique of applying zero-field measurements in small-scale calibration experiments to non-zero field full-scale detectors~\cite{Xe100,ZepIIIb}. This method involves extrapolating the work of Aprile\etal\cite{Aprile2006}, who report the relative light and charge yields versus field for 56.5~keV recoils, well outside the WIMP search windows of experiments to date, which typically cut off at $\sim$20-40 keV~\cite{Xe100,ZepIIIb}. Aprile\etal demonstrated a very slight dependence of yields on field, but only at that energy, and the use of the Dahl points introduces a slight energy dependence, as seen by the deformation of the energy-dependent light yield function in Figure~\ref{NRTwoSquare}, left.

NEST is conservative on this crucial point, which relates to XENON100's energy threshold and consequently to its ability to rule out the low-mass WIMP interpretation of the CoGeNT result~\cite{Xe100,Cogent2013}. In order to reconcile the higher NR light yield measured by Horn\etal and Plante~et~al.~\cite{Horn2011,Plante2011} with the yields measured by Dahl~\cite{Dahl2009}, NEST predicts a steeper drop in light yield ($\sim$0.8 quenching with respect to zero field) than traditionally quoted ($\sim$0.9 even at 4-5~kV/cm) using~\cite{Aprile2006}, as seen in Figure~\ref{NRTwoSquare}, left. This is the case regardless of the energy scale used to normalize the total quanta (light plus charge) at a given NR energy. This is because Dahl provides the ratio of electrons to photons for a fixed sum, constraining their absolute totals. 

NEST takes a combined approach for converting keVee (electron equivalent) into keVnr, defining it using the sum of the charge and light yields, rather than the traditional approach of using light yield alone (because, as explained in~\cite{Szydagis} and ref. therein, it is now well known that energy depositions contribute to both possible channels). However, the choice of conversion factor, Lindhard or otherwise, only changes the absolute NR energy scale, and does not affect our conclusion regarding the field quenching. Even though there is no charge yield at zero field, there is still an electron escape fraction~\cite{Szydagis}. NEST makes the logical assumption of the same reduction in the light plus charge yield, with or without field (the field affecting only the ratio of charge to light), and uses a Thomas-Imel box model recombination factor of 0.19 to match the latest zero-field NR light yield data. However, explaining individual yields measured by Dahl requires a significant step down in light yield when field is present, because from his charge-to-light ratios he gets Thomas-Imel factors between $\sim$0.02-0.04 for fields between 60 and 4,060~V/cm~\cite{Dahl2009}.

An alternative explanation is that the Manzur\etal data~\cite{Manzur2010}, which are $\sim$1-sigma lower than Plante\etal and Horn et al., more accurately represent the light yield at zero field, as argued previously by Collar and McKinsey~\cite{CollarMcK}. In that scenario, not depicted here, employing the same Thomas-Imel parameter as the new one for ER just introduced (0.05), as opposed to the old one ruled out by recent results (0.19), makes NEST simpler, with a similarly smooth transition from zero to non-zero field for NR and ER. The true absolute yield at zero field does not impact the NEST predictions of absolute yield at non-zero field, only the relative yield with respect to zero.

Figure~\ref{NRTwoSquare}, right, depicts charge yield as extracted from Dahl's points, knowing both the sum and the ratio. The light and charge yields are still modeled as anti-correlated, but unlike the ER case, their sum is not a constant in energy, due to loss into a third channel from the Lindhard effect. Existing charge yield data are in large disagreement~\cite{Sorensen2010,Sorensen2010b}. In some cases, the charge yield is extracted from the measured charge-to-light ratio, assuming a light yield curve, which if too high would make the charge yield correspondingly too high~\cite{Xe100Qy,Horn2011}. The curve on the right is provided here to serve as a comparison point for future experiments which plan to measure it more accurately~\cite{AaronTalk,PlanteTalk}. Its points are lower than most existing data, making NEST conservative when applied to low-mass WIMP analyses or to coherent neutrino scattering. Both charge and light yield predictions are provided for the community online in numerical form for ease of use~\cite{NestSite}. Further confrontation with experimental data will be instrumental in confirming Dahl, and the Thomas-Imel recombination model.

As far as we know, only NEST addresses the field dependence of the NR charge yield, as opposed to other works, which depict a generic curve without field dependence, with alternative curves representing different models, but not different fields~\cite{Bezrukov2011,Sorensen2010,Sorensen2010b}. Every data set ever taken~\cite{Manzur2010,Sorensen2010,Sorensen2010b,Horn2011,Xe100Qy} agrees at least qualitatively with the NEST curves to the extent that they all show the charge yield per unit energy increasing with decreasing energy, to the lowest energy reached in data. This is explained by a lowering of recombination probability with decreasing $dE/dx$, as NR track length diminishes (Figure 5 in~\cite{Aprile2006}).

In contrast, the charge yield models presented by Collar (see black lines in Figure 2 of~\cite{Collar2011}), when re-plotted as electrons per unit energy versus energy, as opposed to total electrons versus energy, show a trend opposite to that of world data. For them to be correct, different experiments using unique approaches to determine the NR energy, such as neutron angle of scatter~\cite{Manzur2010}, light alone (e.g.~\cite{Xe100Qy}), and charge plus light~\cite{Dahl2009,Sorensen2011}, must all contain an incorrect determination of the recoil energy. However, there are reasons to believe the charge yield may be lower than often reported (in our case, the Hitachi-corrected Lindhard factor). The NEST charge and light figures are limited to above 1~keV because of uncertainty in simulating non-monotonicity at low recoil energies due to atomic effects~\cite{LindPriv}. However, NEST is capable of providing a prediction at any energy, down to $\sim$1 electron ionized on average at $\sim$300~eV, consistent with the work of Sorensen~\cite{SorensenPriv}.

\section{Conclusion} %7
\label{sec:conclusion} %7

We have presented an update on the NEST simulation package, which has extended its comprehensiveness to include NR charge yield, including its dependence on the magnitude of the electric field, and the dependence of the light yield on field. Even though NEST does not derive everything strictly from first principles, it has a firm grasp of the microphysics, providing a lens with which to view all data, for any particle type, energy, and field. While it does not track individual atoms or excimers due to the prohibitive computational effort required, it does consider the excitation, ionization, and recombination physics. We expect that NEST will continue to positively impact all developments of xenon-based detectors.

\acknowledgments

This work was supported by U.S. Department of Energy grant DE-FG02-91ER40674 at the University of California, Davis, as well as supported by DOE grant DE-NA0000979, which funds the seven universities involved in the Nuclear Science and Security Consortium. We also would like to thank the Extreme Science and Engineering Environment (XSEDE) for providing supercomputing resources to the NEST project.

\bibliographystyle{JHEP}
\bibliography{NESTPaper}
\end{document}